\documentclass[prb,showpacs,twocolumn,amsmath]{revtex4}
\usepackage{latexsym}

\newcommand{\be}{\begin{equation}}

\newcommand{\ee}{\end{equation}}

\newcommand{\giro}[1]{\stackrel{\circ}{#1}}

\begin{document}

\title{ Scissors Modes: The elusive  breathing overtone}




\author{Keisuke Hatada$^{1,2}$, Kuniko Hayakawa$^{2}$, Fabrizio Palumbo$^{2}$}

\affiliation{$^1$ Scienze e Tecnologie, Universita' di Camerino,
Via Madonna delle Carceri 9, 62032 Camerino, Italy,\\
$^2$INFN Laboratori Nazionali di Frascati, 00044 Frascati, Italy}


\pacs{24.30.Cz,24.30Gd,21.10.Re, 21.60.Ev}

\date{\today}

\begin{abstract}
The Two-Rotor Model predicts two levels  above the Scissors Modes with degenerate  intrinsic energy. They have $J^{\pi}= 0^+,2^+$ and are referred to as overtones.  Their  energy is  below threshold for nucleon emission, which should make them observable. The $J^{\pi}=0^+$ overtone,  that has the structure of an isovector breathing mode, has vanishing $E0$ amplitude so that cannot be directly excited, but it could be reached in the decay of the $J=2^+$ overtone. We discuss such a process and evaluate the $B(E2)$ strength, which, however, turns out to be very small.
\end{abstract}

\maketitle

The Two-Rotor Model, which led to the prediction of the Scissors Modes~[\onlinecite{LoIu}],  has approximately  the spectrum of a planar harmonic oscillator, with some constraints on the states. 
There are two Scissors Modes with $J^{\pi} =1^+,2^+$. They differ by the rotational energy of the nucleus, but have the same intrinsic energy of the order of $ 3 \, MeV$ in the rare earth region~[\onlinecite{Bohl,Ende}].  The first  levels above the Scissors Modes, referred to as overtones,  have $J^{\pi}=0^+,2^+$ and equal intrinsic  energy of the order of $ 6 \, MeV$. The nature of the collective motion of the $0^+$ was known since the beginning, it is a kind of isovector breathing mode. The  $2^+$ was considered in Ref.~[\onlinecite{Sun}]  but its structure in the Two-Rotor model   has been determined only recently~[\onlinecite{Hata}]. It is   a superposition of an isovector breathing mode and a Scissors Mode with K=2.

In general one cannot expect that the higher lying excitations of semiclassical models are  realized in Nature, the more so for the Two-Rotor Model, because of the modest collectivity and substantial fragmentation of the Scissors Modes~[\onlinecite{Bohl, Ende}]. In this case, however, the energy of the overtones is  below threshold for nucleon emission,  so that their  widths  should  remain of purely electromagnetic nature and small enough for these states to be observable. Even so there was not much hope to excite these states.  Indeed the $E0$ amplitude   for the $0^+$ overtone vanishes, while the $E2$ amplitude for the $2^+$ overtone was expected to be proportional to   $\theta_0$, the amplitude of the zero point oscillation,  which in the rare earth region is of order $10^{-1}$. 

We remind that excitation amplitudes in the Two-Rotor Model    are proportional to powers of   $\theta_0$. Now $ B(M1)\uparrow_{scissors}  \, \sim 1/ \theta_0^2$, but $B(E2)\uparrow_{scissors} \, \sim \theta_0^2$, the strength which was expected for the $2^+$ overtone.
 All the other methods,  the schematic  Random Phase Approximation~[\onlinecite{Suzu}], the Interacting Boson model~[\onlinecite{Iach}], the sum rule method~[\onlinecite{Lipp}] and a geometrical model~[\onlinecite{Roho}] give similar results for the  Scissors Modes.   
 
Recently, nevertheless, we decided~[\onlinecite{Hata}]  to study  the structure and  the electromagnetic strengths of the $2^+$ overtone 
 in the Two-Rotor Model. We disregarded  the $0^+$ one, because we did not know how to excite it in the framework of the Two-Rotor Model. 
We found~  that the dominant term of the $B(E2)$ is of  zero order in the expansion with respect to $\theta_0$, and precisely
 \be
 B \left( E2, 0^+ \rightarrow 2_1^+\right) = {3 \over   64 } e^2 Q_{20}^2  \label{BE2} 
 \ee
 where the $2^+$ overtone is denoted $2_1^+$ to distinguish it from the $2^+$ Scissors Mode (see the Table). This value must be compared
 with
 \be
 B(E2)\uparrow_{scissors}= B(E2, 0^+ \rightarrow 2^+)= 3 \theta_0^2e^2 Q_{20}^2  \label{BE2sciss}\,.
 \ee
 The numerical factor in Eq.(\ref{BE2}) largely compensates for the presence of the small factor $\theta_0^2$ in Eq.(\ref{BE2sciss}), but the strength of the overtone remains almost a factor of 2 larger than that of the $J=2^+$ Scissors Mode in the rare earth region.
We expect that this result should be confirmed at least in the Interacting Boson Model, whose Hamiltonian  in the semiclassical approximation reproduces ~[\onlinecite{Diep}]  that of the Two-Rotor Model, with one qualification to be mentioned later.

 Concerning experimental observation, as far as we know, only recently there have been investigations  aimed at getting systematic information~[\onlinecite{Chyz}] on Scissors Modes
 in Gd isotopes up to about 6 MeV. These experiments are very interesting not only because they show that the M1 strength from Scissors modes is required to explain their results, but also because they suggest the existence of Scissors Modes in excited states. They do not give, however, any evidence of our overtones. In our opinion this cannot   be taken as a conclusive proof that the $2^+$ state does not exist. The cited works, in fact, study neutron-photon reactions that cannot be analyzed in the framework of  the Two-Rotor Model, so that we do not know what  is the probability of excitation of the $2^+$ overtone after neutron capture. Smallness of this probability might explain the experimental result. 
 
 In order to assess this crucial point,  from the theoretical point of view we need  a microscopic model, and probably the most suitable one is the Fermion-Boson Interacting model, which, as already reminded, in the boson sector in the semiclassical limit exactly reproduces[\onlinecite{Diep}] the Two-Rotor Model Hamiltonian. From the experimental side we need experiments with electromagnetic interactions.
 
If the $2_1^+$ exists, however, it could provide a way for a direct observation of the $0_1^+$ overtone because  also the  
$B(E2, 2_1^+ \rightarrow 0_1^+)$  is of zero order in $\theta_0$. This process   should occur via  radiation of   photons of  energy 
\be
 E= \frac{3 \hbar^2}{{\mathcal I}} \label{E}
 \ee
where ${\mathcal I}$ is the moment of inertia of the nucleus.  It is then interesting  to study this decay.  Firstly,  in order 
to  establish whether   the isovector breathing mode can actually be seen. Needless to say, its observation in spite of the modest collectivity and substantial fragmentation of the Scissors Modes, would be of much consequence for the picture of Scissors Modes provided by the Two-Rotor Model, the more so in view of the very specific structure of the $2_1^+$ overtone.  Secondly, existence, but also absence, of such a decay might  be relevant to the identification of the $2_1^+$ overtone.  In any case the evaluation of the $B(E2)$ for such a process is necessary for the knowledge  of the total electromagnetic width of this level. A further reason of interest is related to the other electrically charged systems for which Scissors Modes have been predicted~[\onlinecite{Lipp1}].
 
 In the present paper we report the   evaluation of the $B(E2)$ for the decay of the $2_1^+$ overtone into the breathing one, and the $M1$ strength for the subsequent decay of the breathing mode into the  $1^+$ Scissors Mode.  This second strength turns out to be equal to the strength for exciting the $1^+$ Scissors mode from the ground state. To our surprise, instead,   the first one is negligible  due to a cancellation that appears totally accidental. The breathing overtone remains therefore elusive in the framework of the Two-Rotor model.

 In order to make the paper a minimum self contained we report the relevant features  of the Two-Rotor Model. Its  Hamiltonian is 
 \be
 H={ {\vec J}^2 \over 2  {\mathcal I} } +H_{intr} 
 \ee
where the  first term is the total rotational energy of the nucleus, ${\vec J} $ being the total angular momentum and 
 the second one the intrinsic  Hamiltonian which, in the  slightly modified form of Ref.~[\onlinecite{DeFr}], neglecting terms of order $\theta_0$   is

 \begin{eqnarray}
H_{intr}&=& { 1 \over 2  {\mathcal I} } \Bigg[ - { d^2 \over d \theta^2 } -\left( 2+  \cot^2 ( 2\theta)\right)
\nonumber\\
&+& \cot^2 \theta \, J_{\zeta}^2 + \tan^2 \theta  J_{\eta}^2  \Bigg] +V(\theta)\,.
\end{eqnarray}
 $\theta$ is half the angle between the proton neutron axes, $J_{\xi}, J{\eta}$ and $J_{\zeta}$ the angular momentum components along the  axes of the body-fixed  frame and $V$ the proton-neutron interaction potential.  
The range of $\theta$ can be separated into two regions
\be
s_I= s(\theta) s\left({\pi \over 4}-\theta\right), \,\, s_{II}= s\left({\pi \over 2}-\theta \right) \,s\left(\theta -  {\pi \over 4} \right), 
\ee
where $s(x)$ is the step function: $ s(x)=1, x>0$ and zero otherwise. They are obtained from each other by  the reflection of $\theta$ with respect to ${\pi / 2}$. It is convenient to introduce the notation
\be
R_{\theta}f(\theta) = \giro{f}(\theta)
\ee
where
\be
\giro{f}(\theta) = f \left( {\pi \over 2} - \theta \right)\,,
\ee
so that $ \giro{s}_I = s_{II} $.
 We assume $\giro{V}=V$, as appropriate to the geometry of the system.
 Since we know that the angle between the neutron-proton axes is 
 very small  we can assume for the potential a quadratic approximation
\be
V= {1\over 2} C \, \theta_0^2 \, x^2 s_I + {1\over 2} C \,\theta_0^2 \, y^2 s_{II}
\ee
where
\be
\theta_0 = ({\mathcal I}C)^{-{ 1\over 4}}\,, \,\,\,x= {\theta \over \theta_0}\,,  \,\,\, y= { {\pi \over 2} - \theta \over \theta_0}\,.
\ee
The intrinsic hamiltonian is then invariant with respect to the transformation
\be
R= R_{\xi} \left({\pi \over 2} \right) R_{\theta} 
\ee
where $R_{\xi} $ is the rotation operator around the $\xi$-axis,
so that we can study the eigenvalue equation separately in the regions I, II.

The qualification mentioned at the beginning relative to the semiclassical limit of the Interacting Boson Model, is that this limit has been performed~[\onlinecite{Diep}] only for the region I. We certainly do not expect any surprise for   the region II, but the symmetries of the Hamiltonian with the consequent constraints on the states remain to be worked out in this model.

The eigenfunctions and eigenvalues of $H_{intr}$ in region I are~[\onlinecite{DeFr}]
\begin{eqnarray}
\varphi_{Kn}(x) &=& \sqrt{ { n! \over (n+K)! \, \theta_0}} \, x^{K+{1 \over2}} \, L_n^K\left(x^2 \right) e^{-{ 1\over 2} x^2}
\\
\epsilon_{Kn}&=& \omega (2n +K +1)\,, \,\,\omega = \sqrt{C \over {\mathcal I}}
\end{eqnarray}
where the $ L_n^K $ are Laguerre polynomials and the wave functions $\varphi$ are
normalized according to
\be
\int_0^{\infty} dx \, \left(\varphi_{Kn}(x)\right)^2 = { 1\over 2}\,.
\ee
 Enforcing the symmetries of the Hamiltonian and of the system (separate rotations of the proton and neutron bodies through $\pi$ around the $\xi$-axis),  one finds the following results~[\onlinecite{LoIu}],[\onlinecite{DeFr}]. Restricting ourselves to states of positive parity we found
 \be
\Psi_{JM \sigma} = \sum_{K \ge 0}{\mathcal F}^J_{MK}(\alpha, \beta, \gamma) \Phi_{JK\sigma}(\theta) \label{Intr}
\ee
where
\be
{\mathcal F}^J_{MK}= \sqrt{{2J+1}\over 16( 1 +\delta_{K0}) \pi^2 } \left( {\mathcal D}^J_{MK} +(-1)^J{\mathcal D}^J_{M-K}   \right). 
\ee
$J,M$ are the nucleus angular momentum and its component on the $z$-axis of the laboratory system, and $\sigma$ labels the different states with the same $J$. When it is zero it will be omitted. We will make an  assignment of this quantum number  different from previous papers, in which we ignored the $J=0^+$ overtone.
We impose the normalization
\be
\int_0^{2\pi} d\alpha \int_0^{\pi}d \beta \sin \beta \int_0^{2 \pi }d\gamma \int_0^{{\pi \over 2}}d \theta \, |\Psi_{IM\sigma}|^2 =1\,.
\ee
Secondly the expressions of the intrinsic functions $\varphi$ in regions I and II are related according to   \begin{eqnarray}
\Phi_{00}&=& \varphi_{00} \, s_I +\giro {\varphi}_{00}  s_{II}
\nonumber\\
\Phi_{11}&=&\Phi_{210} =  \varphi_{10} \, s_I - \giro{\varphi}_{10}  s_{II}
\nonumber\\
\Phi_{001}&=& \varphi_{01} \, s_I +\giro {\varphi}_{01}  s_{II} 
\nonumber\\
\Phi_{201}&=&{ 1\over \sqrt 2} \left[ \varphi_{01} \, s_I - { 1\over 2} \left( \sqrt{3} \giro{\varphi}_{20} + \giro{\varphi}_{01} \right) s_{II}  \right] 
\nonumber\\
\Phi_{221}&=& { 1\over \sqrt 2}\left[   \varphi_{20} \, s_I + {1\over 2}  \left(  \giro{\varphi}_{20} - \sqrt{ 3} \giro{\varphi}_{01} \right) s_{II}  \right] \,.
\end{eqnarray}
Even if the nucleus in its ground state has axial symmetry, this symmetry is in general lost in the excited states, so that the component of angular momentum along any body-fixed axis is not conserved, resulting in a superposition of intrinsic states with different $K$ and $n$ quantum numbers. 
All the above states however, with the exception of the $J=2^+$ overtone, are pure $K$ and $n$ states. The  ground state $0^+$ has quantum numbers  $J=\sigma=K=n=0$, the Scissors Modes $1^+, 2^+$ have quantum numbers $\sigma=0,  K=1, n=0$ and $J=1, 2$ respectively, the breathing overtone,  $0_1^+$,  $J=0, \sigma=1, K=0, n=1$.  The other overtone $2_1^+$ has $J=2, \sigma=1$ and coupled intrinsic components with $K=2,n=0$ and $K=0,n=1$. 
 
 %
\begin{table} 
   
\begin{tabular}{|l|cccc|} 
\hline
quantum numbers  & $J^{\pi}_\sigma$ & $K$ & $n$ & energy \\
\hline
  ground state     & 0$^+$  & 0 & 0  &0\\
\hline
  Scissors Modes  & 1$^+$ & 1 & 0  &$\hbar \omega + \hbar^2/ { \mathcal I}$\\[-1.0mm]
                                 & 2$^+$  & 1 & 0 &$\hbar \omega+ 3\hbar^2 / {\mathcal I}$\\
\hline
  breathing overtone & 0$^+_1$  & 0 & 1 &$2\hbar \omega$\\
\hline
 $2^+$ overtone  & 2$^+_1$  &  2 & 0  &$ 2\hbar \omega+3\hbar^2/ {\mathcal I}$\\[-1.0mm]
          &  &  0 & 1& \\
\hline
\end{tabular}

 \caption{Quantum numbers of the positive parity states of the Two-Rotor Model.}
\end{table}
%

The collective  motion of the $0^+$  overtone is a kind of isovector breathing mode. The $2^+$ overtone is a superposition of the breathing mode  in region I (the state $\varphi_{01}$),  and of  the state $\varphi_{20}$, which is a relative rotation of the neutron-proton axes as in the Scissors Mode but with angular momentum $K=2$. Such mixing   is determined by the different form that the intrinsic Hamiltonian takes in regions I and II. Because of it the $2^+$ overtone  might  be called scissors-breathing mode. All these states with their quantum numbers are reported in the Table.

We already mentioned that  the  spectrum of the intrinsic part of the Two-Rotor Model is identical to that of the planar harmonic oscillator with some constraints due to various symmetries. These constraints  reduce the degeneracy of  the first  excited states of the planar harmonic oscillator from 2 to 1 (the Scissors mode) and of the higher levels   form 3 to 2 (the 2 overtones).

Notice that the normalization of the $\Phi$ in Eq.(\ref{Intr}) is different from that in Ref.~[\onlinecite{DeFr}].

As we said at the beginning,  the  quadrupole operator  to  zero order~[\onlinecite{Hata}]   in  $\theta_0$ is 
\begin{eqnarray}
M(E2,\mu) &=& e \, Q_{20}  \left[  {\mathcal D}^2_{\mu 0}  \left(s_I - { 1\over 2} \, s_{II}  \right) \right.
\nonumber\\
&&\left. +  {1\over 2} \sqrt {3 \over 2} \left( {\mathcal D}^2_{\mu 2} + {\mathcal D}^2_{\mu -2}   \right) s_{II} \right]
\end{eqnarray}
where $e \, Q_{20}$ is the quadrupole moment in the intrinsic frame.
We  then see that while to zero order in $\theta_0$ we  cannot excite the breathing overtone  from the ground state,  we could reach it by   the decay of the $J=2^+$ overtone  with the amplitude
\begin{eqnarray*}
\langle \Psi_{2M1}| M(E2, \mu)| \Psi_{001}\rangle &=& { 1 \over 2 \sqrt 10} \, e \, Q_{20} C_{002\mu}^{2M}
\\
&\times&
\left( \langle \varphi_{01}|\varphi_{01} \rangle+ \sqrt 3 \langle \phi_{20}|\phi_{01} \rangle \right)
\end{eqnarray*}

where $C^{2M}_{002\mu}$ is a Clebsch-Gordan coefficient. Because
\be
\langle \varphi_{01}|\varphi_{01} \rangle= -\sqrt 2 \langle \phi_{20}|\phi_{01} \rangle = \frac{1}{2}
\ee
 the contributions of the $K=0$ and $K=2$ components of the $J=2^+$ overtone  almost cancel out with each other, at variance with  the 
 $ B(E2, 0^+ \rightarrow 2_1^+) $ that is  entirely due  to its $K=0$ component, namely its breathing component~[\onlinecite{Hata}].  Such cancellation appears completely accidental, but it overcompensates the absence of the small factor $\theta_0^2$, so that 
  the transition strength
 \be
B(E2, 2_1^+ \rightarrow  0_1^+) = { 1 \over 32 }\left( 1 - \sqrt { \frac{3}{2}} \right)^2 \, e^2 \, Q_{20}^2 
\ee
results negligibly small.

In conclusion the  electromagnetic width of the  $2_1^+$ overtone  is entirely  due to its $M1$ decay to the Scissors Mode $1^+$ and to its $E2$ decay to the ground state~[\onlinecite{Hata}], and the breathing overtone cannot be reached through its decay.

For the case in which the breathing mode could be excited by other means or appear in intermediate states, we report its $B(M1)$ strength for decaying into the $1^+$ Scissors Mode. The expression of the magnetic dipole operator is~[\onlinecite{Hata}]

\begin{eqnarray} 
 M(M1, \nu) &=&- { 1 \over \sqrt 2}\left( {\mathcal D}^1_{\nu 1}- {\mathcal D}^1_{\nu -1}  \right) {\mathcal M}
 \nonumber\\
 &\times &
\left( s_I - s_{II} \right) \left({d \over d \theta} - { 1 \over 2 \theta} \right)\,,
\end{eqnarray}
where
\be
{\mathcal M} =  i \sqrt {3 \over 16 \pi}  \left( g_p - g_n \right) { e \over 2m_p} \,,
\ee
$g_n,g_p$ being the orbital gyromagnetic factors of neutrons, protons respectively and $m_p$ is the proton mass. 

By a straightforward calculation we get
\begin{eqnarray}
\langle \Psi_{001}|M(M1,\nu) | \Psi_{1M }\rangle &=& 2 \sqrt 3 \, {\mathcal M} \langle \phi_{01}|\nabla_{\theta}|\phi_{10}\rangle
\nonumber\\
&\times&C_{1M1\nu}^{00}C_{1-111}^{00}\,.
\end{eqnarray}
Since
\be
\langle \phi_{01}|\nabla_{\theta}|\phi_{10}\rangle = \frac{1}{2 \theta_0}
\ee
we find
\be
B(M1, 0_1^+ \rightarrow 1^+)= \frac{1}{\theta_0^2}\, |{\mathcal M}|^2\,,
\ee
which is equal to   the $B(M1)$ strength for excitation of the $1^+$ Scissors mode from the ground state.


 \end{document}